\documentclass[12pt]{article}
\hoffset = - 0.5 truecm \voffset=-1.4 truecm
\def\beq{\begin{equation}}   \def\eeq{\end{equation}}
\def\bea{\begin{eqnarray}}  \def\eea{\end{eqnarray}} 
 \def\beeq{\begin{eqnarray}}
\def\eeeq{\end{eqnarray}}
\def\lsim{\raise0.3ex\hbox{$<$\kern-0.75em\raise-1.1ex\hbox{$\sim$}}}
\def\gsim{\raise0.3ex\hbox{$>$\kern-0.75em\raise-1.1ex\hbox{$\sim$}}}
\usepackage{wrapfig,rotating}
\usepackage{graphicx}
\usepackage{epsfig}

\renewcommand{\theequation}{\thesection.\arabic{equation}}
\newcounter{hran} \renewcommand{\thehran}{\thesection.\arabic{hran}}

\def\bmini{\setcounter{hran}{\value{equation}}
         \refstepcounter{hran}\setcounter{equation}{0}
         \renewcommand{\theequation}{\thehran\alph{equation}}\begin{eqnarray}}

\def\bminiG#1{\setcounter{hran}{\value{equation}}
\refstepcounter{hran}\setcounter{equation}{-1}
\renewcommand{\theequation}{\thehran\alph{equation}}
\refstepcounter{equation}\label{#1}\begin{eqnarray}}

\def\emini{\end{eqnarray}\relax\setcounter{equation}{\value{hran}}\renewcommand{\theequation}{\thesection.\arabic{equation}}}

\begin{document}

\begin{titlepage}

\begin{flushright}
LAPTH-1320/09\\
LPT-Orsay 09-21\\
IPPP/09/17\\
DCPT/09/34

\end{flushright}
%\vspace{1.cm}

\begin{center}
{\large \bf 
PHOTON - JET CORRELATIONS \\
AND CONSTRAINTS ON FRAGMENTATION FUNCTIONS}
\par 

%\vskip 3 truemm

\vskip 1 truecm 

{\bf 
%P.~Aurenche$^{(b)}$, 
Z. ~Belghobsi$^{(a)}$, M. ~Fontannaz$^{(b)}$, 
J.-Ph. ~Guillet$^{(c)}$, G.~Heinrich$^{(d)}$, E.~Pilon$^{(c)}$, 
M.~Werlen$^{(c)}$
} 

\vskip 2 truemm

{\it $^{(a)}$ Laboratoire de Physique Th\'eorique, LPTh, Universit\'e de Jijel
BP 98 Ouled Aissa, 18000 Jijel, Algeria}

{\it $^{(b)}$ Laboratoire de Physique Th\'eorique, UMR 8627 du CNRS,\\
Universit\'e Paris XI, B\^atiment 210, 91405 Orsay Cedex, France}\\

\vskip 2 truemm

{\it $^{(c)}$ LAPTH, Universit\'e de Savoie, CNRS,\\ 
BP. 110, F-74941 Annecy-le-Vieux Cedex, France}

\vskip 2 truemm

{\it $^{(d)}$ IPPP, Department of Physics, 
University of Durham, Durham DH1 3LE, UK}

\vskip 7 truemm

\normalsize

\begin{abstract}

We study the production of a large-$p_T$ photon in association with a jet in 
proton-proton collisions. We examine the sensitivity of the jet rapidity 
distribution to the gluon distribution function in the proton. 
We then assess the sensitivity of various photon + jet correlation observables 
to the photon fragmentation functions. We argue that RHIC data on photon-jet 
correlations can be used to constrain the photon fragmentation functions in a 
region which was barely accessible in LEP experiments.

\end{abstract}

\end{center}

\begin{center}
\today
\end{center}

\end{titlepage}

\section{Introduction}\label{cor-intro}

The phenomenology of prompt photons is very rich and interesting, as the photon
on one hand can be considered as a pointlike particle  described by QED, leading
to clean experimental signatures. On the other hand, the photon is also
involved in hadronic phenomena, like the fragmentation of an energetic parton  
into a large-$p_T$ photon and hadronic energy. 

\vspace{0.3cm}

At the LHC, photon+jet final states will be important for jet calibration and
pdf 
studies~\cite{Ball:2007zza,:2008zzk,Aad:2009wy,:2008zzm,{Stockton:2008zz},PerezReale:2008zza}.
Diphotons will play an important role in the search for a Higgs boson with mass below $\sim$ 140
GeV, where the decay into two photons is a very prominent
channel~\cite{Buescher:2005re,Stockli:2005hz}, for which the branching ratio is 
small, $\sim {\cal O}(10^{-3})$, and the signature is provided by a narrow peak 
over a huge background involving various components. Besides the so-called
irreducible background from prompt diphotons, the background called reducible
comes from photon-jet and jet-jet events, with the jet faking a photon in
various possible ways (high $p_{T}$ $\pi^{0}$ or other neutral hadrons, charged
particles inducing the radiation of energetic photons e.g. by bremsstrahlung due
to interactions with innermost layers of the detector, etc.). An accurate
knowledge of the photon+jet rate in particular is required to estimate
and control the reducible background to the Higgs boson search in the diphoton
channel \cite{Fang:2008zzc}. 
In addition, highly energetic photons are important signatures for
various scenarios of physics beyond the Standard Model. Therefore, issues like
controlling photon isolation or further constraining the parton-to-photon
fragmentation functions are of major importance, and data from RHIC and the
Tevatron should be exploited as much as possible to this aim.

\vspace{0.2cm}

The production of prompt photons in hadronic collisions may be schematically
seen as originating from either of two mechanisms. In the first one, which may 
be called `direct' (D), the photon behaves as a high $p_{T}$ colourless parton, 
i.e. it takes part in the hard subprocess, and is most likely well 
separated from any hadronic environment. In the other one, which may be called
`fragmentation' (F), the photon behaves hadron-like, i.e. it
results from the collinear fragmentation of a coloured high $p_{T}$ parton. In
the latter case, it is most probably accompanied by hadrons - unless the photon
carries away most of the transverse momentum of the fragmenting parton.

\vspace{0.2cm}

From a technical point of view, (F) emerges from the calculation of the higher
order corrections in the perturbative expansion in powers of the strong
coupling $\alpha_{s}$. 
At higher orders, final state collinear singularities appear in any subprocess 
where a high $p_{T}$ outgoing parton of species $k$ (quark or gluon) undergoes 
a cascade of successive collinear splittings together with the collinear 
emission of a photon. 
The higher order corrections to the cross section can be split into 1) a 
contribution free from these final state collinear singularities, 
to be added to the Born term so as to build (D), and 2) a contribution (F) 
involving these 
singularities together with accompanying large collinear logarithms. 
In (F), the final state collinear singularities and accompanying logarithms can 
be factorised to all orders in $\alpha_s$ from short distance terms according 
to the factorisation theorem, and absorbed into fragmentation functions of 
parton $k$ to a photon $D_{\gamma/k}(z,M_{_F}^{2})$. 
Let us mention however that the splitting of the cross section between (D) and 
(F) is not unique and that the $D_{\gamma/k}(z,M_{_F}^{2})$ depend on the 
arbitrary factorization scheme specifying which non-singular parts are 
factorised together with the collinear singularities; the latter depend in 
particular on some arbitrary fragmentation scale $M_{_F}$. We therefore need to 
define the scheme used. In this article, 
(D) is defined as the Born term plus the fraction of the higher order 
corrections from which 
final state collinear singularities and accompanying collinear logarithms have 
been subtracted according to the ${\overline{\mbox{MS}}}$ factorisation scheme.
(F) is defined as the contribution involving a fragmentation function of any 
parton into a photon defined in the ${\overline{\mbox{MS}}}$ scheme. 
The partonic cross section can thus be written schematically as:

\vspace{0.2cm}

\begin{equation}
{\rm d}\sigma^{\gamma} = {\rm d}\sigma^{(D)}(\mu^{2},M^{2},M_{F}^{2})
+ \sum_{k=q,\bar{q},g} {\rm d}\sigma^{(F)}_{k}(\mu^{2},M^{2},M_{F}^{2}) 
\otimes D_{\gamma/k}(M_{F}^{2})
\label{e1}
\end{equation}

\vspace{0.2cm}

\noindent
where $\mu$, $M$, $M_{F}$ are respectively the (arbitrary) renormalization, 
initial state factorisation and final state fragmentation scales, and
``$\otimes$" stands for a convolution over the fragmentation variable. 
The point-like coupling of the photon to quarks is responsible for the 
well-known anomalous behaviour of $D_{\gamma/k}(z,M_{_F})$, 
roughly as $\alpha_{em}/\alpha_{s}(M_{_F}^{2})$, when 
the fragmentation scale $M_{_F}$, chosen of the order of a hard scale of the
subprocess, is large compared to ${\cal O}$(1~GeV). More generally, while the
$M_{F}$ evolution of these fragmentation functions is given by inhomogeneous
evolution equations whose kernels are computable in perturbative QCD, the $z$
profiles of these fragmentation functions are not fully predictable from 
perturbative QCD. These parton-to-photon fragmentation functions therefore 
either have to be modeled in some way and/or constrained using experimental 
data. 

\vspace{0.2cm}

The fragmentation component represents a fraction of the inclusive prompt 
photon signal which grows with the center-of-mass energy of the collision.
While it remains subleading at fixed target energies, it becomes dominant at
collider energies. On the other hand, most collider experiments -- apart from
the PHENIX experiment at RHIC \cite{Peressounko:2006qs}, but, in particular, 
the TeV collider experiments CDF and D0 at the Tevatron, ATLAS and CMS at the 
LHC --  do {\em not} measure 
inclusive photons, because at these energies the inclusive prompt photon 
signal would be swamped by a large background of secondary photons from 
decays of fast neutral mesons (mainly $\pi^{0}$, as well as $\eta$, etc.). 
Instead these experiments impose isolation criteria on the hadronic final 
states of photon candidate events,
requiring that the photon be not accompanied by more than a prescribed amount 
of hadronic transverse energy in some given cone about the photon. An 
analogous\footnote{The experimental request may also impose a veto on charged 
tracks in the vicinity of the photons. However such vetoes cannot be 
implemented in partonic level calculations: a full description of the 
hadronised final state would be necessary.} criterion can be implemented in 
parton level calculations. 
The isolation cuts do not only suppress the background, they also substantially 
reduce the (F) component. Yet some fraction of the (F) component may 
survive and affect shapes of various tails of distributions, especially for
correlation observables. Of course the sensitivity is even larger when 
loose isolation cuts are applied. 
In this article we wish to stress the interest of
photon-jet correlation observables, in particular in constraining the photon 
fragmentation functions, which requires to go beyond the lowest order. 
Similar studies have been performed in ref. 
\cite{Baer:1989xj}. However 
in these works 
the (F) component was calculated
at Lowest Order (LO) only.

\vspace{0.2cm}

In a previous article \cite{Aurenche:2006vj} we proposed a critical
re-examination of the status of single prompt photon production in hadronic
collisions in the light of recent experiments, which was based on a
Next-to-Leading Order (NLO) calculation of both (D) and (F) provided in the 
form of a partonic Monte-Carlo code, {\tt JETPHOX} \cite{http-jetphox}. 
The present paper aims at supplementing this previous work with a study of 
photon-jet correlations using the same tool, for the 
presentation of which we refer to \cite{Aurenche:2006vj}. 
The article is organised as follows. In section \ref{ctstar}, we 
examine the magnitude of the fragmentation component on the photon-jet angular 
distribution at the 
Tevatron. In section \ref{yjet} we discuss the jet rapidity distribution in 
photon $+$ jet associated production, as a possible way to help constrain 
the uncertainties on the gluon distribution function. In sect. \ref{pho-frag} 
we then 
discuss the potential of photon-jet correlations measured at RHIC without 
isolation as a tool to constrain the photon fragmentation functions. 
Finally, section \ref{conclusions} gathers our conclusions.
A similar study dedicated to fragmentation into hadrons will be discussed in a
future article.

\section{Photon-jet angular distribution}\label{ctstar}

An observable expected to receive a distinctive contribution from the (F) 
component is the photon-jet angular distribution which has been measured by the 
CDF Collaboration \cite{Abe:1993cv,Lamoureux:1995ps} and is defined as follows. 
  
\vspace{0.3cm}

At lowest order (LO), corresponding to $2 \to 2$ kinematics, $\cos \theta^{*}$ 
is the cosine of the angle between the photon direction and the beam axis in 
the center-of-mass system of the partonic subprocess. It also coincides with
$\cos \theta^{*} = \tanh y^{*}$ where 
$y^{*} = ( y_{\gamma} - y_{\rm{jet}})/2$. This angular distribution 
is expected to receive a dominant contribution from the (F) component when 
$\cos \theta^{*}$ becomes close to 1. Indeed, at lowest order, the (D) 
component proceeds via a $t$-channel quark exchange yielding a behaviour 
$\sim 1/(1- \cos \theta^{*})$ for the partonic amplitude squared, whereas the 
(F) component involves also gluon exchange in the $t$-channel, yielding a 
behaviour $\sim 1/(1- \cos \theta^{*})^{2}$. On this ground, one thus expects 
(F) to take over for $\cos \theta^{*}$ values close enough to 1.  
  
\vspace{0.3cm}

\begin{figure}[htb]
\begin{center}
\epsfig{file=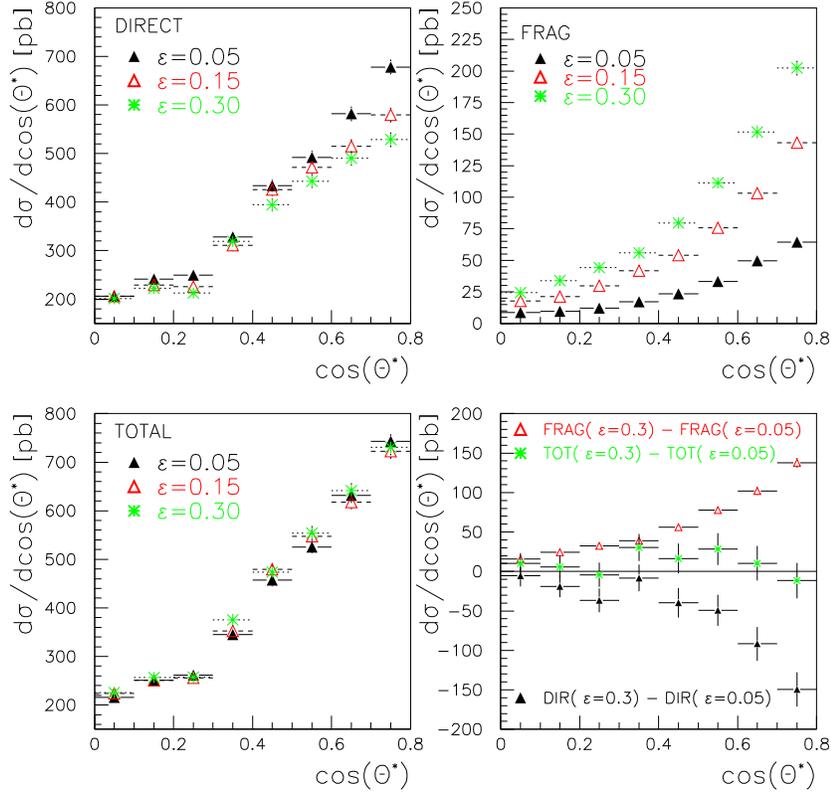, height=12cm}
\caption{\label{cst-three} 
Sensitivity of the distribution of $\cos \theta^{*}$ in photon + jet at NLO to 
the isolation parameter $\epsilon = E_{t \, max} / p_{T}^{\gamma}$.
Top left: Direct (D) contribution only. 
Top right: Fragmentation (F) contribution only. 
Bottom left: total (D) + (F) contribution. 
Bottom right: differences in (D), (F) and total (D)+(F) between $\epsilon =$ 0.3
and 0.05.}
\end{center}
\end{figure}
Depending on the cuts applied, the $\cos \theta^{*}$ dependences coming from 
the partonic transition matrix element squared may be blurred by an extra 
dependence coming through the parton luminosity. Focusing on the direct (D) 
contribution, and parametrizing the LO phase space as 
\begin{equation}\label{ps1}
d(p_{T}^{\gamma})^{2} \, dy_{\gamma} \, dy_{\rm{jet}}
= \frac{1}{2} \, d(p^{*})^{2} \, dy_{B} \, d \cos \theta^{*}
\end{equation}
in terms of the variables $\cos \theta^{*}= \tanh y^{*}$,
$y_{B} = (y_{\gamma} + y_{\rm{jet}})/2$ and $
p^{*} = p_{T}^{\gamma} \, \cosh y^{*}$, 
the LO distribution in $\cos \theta^{*}$ reads:
\begin{equation}\label{ect1} 
\frac{d \, \sigma}{d \cos \theta^{*}} = \sum_{i,j}
\int dy_{B} \, dp^{*} \, G_{i/P}(x_{i}) \, G_{j/\bar{P}}(x_{j})
\frac{d \, \hat{\sigma}_{ij}}{d \cos \theta^{*} \, dy_{B} \, dp^{*}}
\end{equation}
with
\begin{equation}\label{xij}
x_{i,j} = \frac{2 p^{*}}{\sqrt{S}} \, e^{\pm y_{B}}
\end{equation}
In particular, one of the two parton distribution functions (pdfs) involved in 
the distribution in $\cos \theta^{*}$ has an argument $x$ which grows with 
$y^{*}$ i.e. with $\cos \theta^{*}$ at fixed $p_{T}^{\gamma}$, so that this 
pdf decreases (towards zero) if $\cos \theta^{*}$ increases (towards one). 
In absence of extra cuts, this decrease actually takes over the growth of the 
partonic cross section with growing $\cos \theta^{*}$ over the whole range.
At LO, this can be neutralised by imposing cuts on $y_{B}$ and $p^{*}$ 
independent from $y^{*}$ \cite{Abe:1993cv,Lamoureux:1995ps}, so that the 
integration over $y_{B}$ and $p^{*}$ in eq. (\ref{ect1}) yields 
$\cos \theta^{*}$ independent factors. Similar procedure and conclusion hold 
for the fragmentation (F) component, at least in the absence of isolation.
  
\vspace{0.3cm}

Beyond LO the definition of $\cos \theta^{*}$ has to be extended.
This extension is not unique, and various definitions 
can be found in the literature. Here we take\footnote{An alternative 
possibility is the one used in \cite{Abe:1993cv}, which, in short, combines 
the several jets of a multijet final state into one so-called superjet 
recoiling against the photon, in order to stick to a $2 \to 2$ 
kinematics as close as possible. See \cite{Abe:1993cv} for more details.}

\begin{equation}\label{costhstar-blo1}
\cos \theta^{*} = \tanh y^{*}
\end{equation}
where 
\begin{equation}\label{ystar}
y^{*} = \frac{1}{2} \left( y_{\gamma} - y_{\rm{leading \, jet}} \right)\;,
\end{equation}
$y_{\gamma} - y_{\rm{leading \, jet}}$ being the difference\footnote{Note
that the definition of $\cos \theta^{*}$ given by (\ref{costhstar-blo1}),
(\ref{ystar}) refers to a quantity which is invariant under longitudinal boosts 
along the beam axis.} between the rapidity of the photon and the rapidity of 
the {\em leading} jet, i.e. the jet of highest transverse energy. 
Furthermore, the higher order contributions to the angular distribution involve 
an extra convolution smearing over kinematical configurations, 
so that the interpretation of this distribution beyond LO is less 
transparent.

\vspace{0.3cm}

Besides, measurements at colliders most often involve isolated 
photons, in which case the (F) component is quite reduced. Namely, when 
the hadronic transverse energy accompanying the photon is required 
to be smaller than $E_{T \, max}$, 
the (F) contribution is roughly proportional to 
$(1-z_{c}) \simeq E_{T \, max}/p_{T}^{\gamma}$, the width of the support 
$[ \, z_{c} \, , \, 1 \, ]$ of the convolution with the photon fragmentation
functions. If $E_{T \, max}$ is chosen such that this ratio is always small, 
the dominance of the $t$-channel gluon exchange from the (F) component is
never effective; yet one might still expect a sizable distortion of the angular 
distribution for $\cos \theta^{*}$ close enough to 1. We note also that if 
$E_{T \, max}$ does not scale with $p_{T}^{\gamma}$, the isolation constraint
implied on the fragmentation variable $z$, $z \geq z_{c}$ with
\begin{eqnarray}
1 - z_{c} 
& \simeq & 
\frac{E_{T \, max}}{p^{*}} \frac{1}{(1 - \cos^{2} \theta^{*})^{1/2}} 
\label{isolcosth}
\end{eqnarray}
induces, through the convolution over $z$, an extra dependence on 
$\cos \theta^{*}$ which contorts (amplifies somewhat) the growth of the (F) 
contribution provided by the partonic transition matrix elements alone.

\vspace{0.3cm}
 
In a preliminary study \cite{Lamoureux:1995ps} subsequent to the analysis 
published in \cite{Abe:1993cv} and based on a data set with larger
statistics and extended towards lower
values of $p_{T}^{\gamma}$, the CDF Collaboration found a 
discrepancy between the measured $\cos \theta^{*}$ distribution and the 
theoretical prediction of \cite{Baer:1989xj}. 
This subsequent preliminary CDF analysis concluded that extra dijet-like 
contributions involving $t$-channel gluon exchange would be necessary to bridge 
the gap, and that these extra contributions might come from NLO contributions 
to the (F) component.
Since the prediction of \cite{Baer:1989xj}
involves an account of (F) at LO only, we have revisited this observable and 
computed the effects of accounting for (F) at NLO. 
The CDF Collaboration adopted a procedure to patch together the contributions 
from data corresponding to two regions that were distinct in $p^{*}$ and 
$y_{B}$ though overlapping in $\cos \theta^{*}$, in order to maximise the 
range in $\cos \theta^{*}$ displayed on one and the same plot. 
In particular the distribution was normalized to 1 in the 
%three bins farthest
bin farthest 
from $\cos \theta^{*} = 1$, and the data sets from the two regions were 
normalized to each other in one overlapping bin in $\cos \theta^{*}$. 
We understand this approximate procedure to have been dictated by the use of 
limited statistics and precision of Run I data but we did not follow the same 
procedure in our study for several reasons. 
Firstly, the normalization to 1 in the bin farthest from 
$\cos \theta^{*} = 1$ aims at getting rid of the numerical factor coming from 
the integration of partonic luminosity. This is fine as long as only one 
partonic subprocess contributes - or at least, when one yields a much greater 
contribution than all the others. However, in the present case, the 
distribution is of the form
\begin{equation}\label{weight}  
\frac{d \sigma}{d \cos \theta^{*}} = 
\sum_{s} {\cal L}^{(s)} \frac{d \hat{\sigma}^{(s)}}{d \cos \theta^{*}}
\nonumber
\end{equation}
i.e. a linear combination of contributions coming from several subprocesses $s$.
In particular, 
considering (D) only, $gq$ (or $g\bar{q}$) initiated and $q \bar{q}$ initiated 
processes which contribute at LO have distinct functional dependences on 
$\cos \theta^{*}$.
Secondly, the integrated partonic luminosity factors ${\cal L}^{(s)}$ depend not
only on the subprocess $s$ but also on the integration regions in phase space. 
The relative weights of the subprocesses from the contributions (D) and (F) 
thus differ in the two regions defined by CDF as ``1" and ``2". 
The normalization enforced by the matching procedure of CDF is not harmless on 
the impact of the NLO correction to the (F) contribution, and might 
{\em reduce} the impact of this correction.
\vspace{0.3cm}

Therefore we did not stick to the CDF study. 
We focused on a study of the magnitude of the (F) contribution, without 
making a direct comparison with the CDF Run I data.
Our study has been made for $\sqrt{S} = 1.96$ TeV, with the following
definitions and kinematic cuts: $|y_{\gamma}| \leq 0.9$, $p_{T} \geq 30$ GeV, 
$p_{T}^{leading \, jet} \geq 25$ GeV. The jets were defined according to the D0 
midpoint algorithm \cite{jetD0}, with cone aperture $R_{c} = 0.7$. 
The photon isolation required that in a cone of aperture $R = 0.4$ in rapidity 
and azimutal angle around the photon direction, the fraction of maximal 
hadronic transverse energy $E_{T \, max}/p_{T}^{\gamma}$ be less than a 
prescribed value $\epsilon$, which we varied from 0.05 to 0.3. The further cut 
45 GeV $\leq p^{*} \leq 55$ GeV was imposed. The pdf set CTEQ 6.1 was used
together with the BFG set II for the fragmentation functions, with the scale 
choice $\mu = M_{F} = M = p_{T}^{\gamma}/2$.
  
\vspace{0.3cm}

We have considered three ingredients which may affect the size of the
contribution  of the (F) component. One is the account for the NLO corrections
to the many subprocesses; another one concerns the uncertainty on the
fragmentation functions; yet another one deals with a possible mismatch between
the implementation of isolation at the partonic vs. hadronic level.

Let us first consider the impact of the NLO corrections to the (F) component.
For the standard scale choice $\mu = M_{F} = M = p_{T}^{\gamma}/2$, the effect 
is to multiply the component (F) at LO by about a factor two. From 
the top-right fig. \ref{cst-three}, the impact on the total (D) + (F) both at 
NLO amounts to an increase by 4 \% in the upper $\cos \theta^{*}$ 
range w.r.t. (D) at NLO + (F) only at LO for $\epsilon = 0.05$.  

Is it possible to increase the (F) contribution by modifying the fragmentation 
functions? When a stringent isolation cut is required on the photon candidates 
as in the CDF experiment, the (F) contribution involves the photon 
fragmentation function at $z \geq z_c$ i.e. rather close to 1. 
In this region the fragmentation functions are dominated by their so-called 
anomalous parts predicted by perturbative QCD. Their poorly known non 
perturbative parts, which would be the only adjustable ingredients, play no 
role: thus the (F) contribution to the $\cos \theta^{*}$ distribution is 
rather tightly constrained.   

We have tackled the issue of the account of isolation at the partonic vs. hadronic
level by varying the value of the isolation parameter $\epsilon$ from 0.05 to
0.3. As already mentioned in table 2 of \cite{cfgp} in the case of
the inclusive cross section and as can be seen on fig. \ref{cst-three}, 
the separate (D) and (F) contributions do depend strongly on $\epsilon$ at NLO; 
yet strong cancellations turn out to occur between (D) and (F) so that the 
total (D) + (F) depends on $\epsilon$ only very mildly, at least 
as long as the infrared sensitive term $\alpha_{s} \ln \epsilon$ in (D) does 
not become large - otherwise the fixed order prediction becomes unreliable. 
Therefore, changing the calorimetric isolation parameter by as much as a factor 
of six does not modify the total contribution to the $\cos \theta^{*}$ 
distribution significantly. 

To summarize, the results of our calculations show 
that the idea of playing with the fragmentation component 
as suggested in the CDF analysis turns out to be ineffective in the conditions 
which we have considered. It would be worthwhile to perform a quantitative
analysis of the much larger statistics data set gathered in Run II, without 
relying on the questionable matching procedure used in the CDF Run I analysis.

\vspace{0.3cm}

Let us recall that, beyond the isolation requirements and the refined analysis 
to improve background rejection, the CDF measurement of isolated photons 
required the statistical subtraction of a contamination of photon candidates 
coming from neutral hadrons - mainly $\pi^{0}$, plus $\eta$, $\rho^{0}$, 
$\omega$ etc. The background was removed statistically by exploitation of the
expected difference in the resulting shower profiles  
or by the different conversion probabilities  
using Monte Carlo simulations \cite{Abe:1993qb}. 
The latter relied on hadronisation models suited to describe the bulk of 
hadronisation. On the other hand, the {\it very small} fraction of hadronic 
events which pass the isolation cuts corresponds to the tail of fragmentation 
at large $z$ which is not constrained by the data.
These background events yield namely dijet-type contributions involving 
$t$-channel gluon exchange, which might explain part of the discrepancy
observed, and the distribution in $\cos \theta^{*}$ at $\cos \theta^{*} \to 1$ 
might provide an enhanced sensitivity to this contamination w.r.t. other
prompt photon observables.

\vspace{0.3cm}

Let us add a comment on the comparison with the situation for the photon-photon
azimuthal angle distribution in photon pair production \cite{Acosta:2004sn}. 
In the latter case, some of the higher order contributions involving one direct photon
and one photon from fragmentation provided a collinear logarithmic enhancement 
at low azimuthal angle, when a hard jet recoils against the photon pair at low 
relative angle. No such phenomenon occurs in the photon-jet case, since the 
jet considered in that observable is always the leading jet of the event which 
roughly recoils against the photon.

\section{Jet rapidity distribution}\label{yjet}

Despite years of intense work, a proper understanding of the uncertainties on
the gluon distribution function is still lacking \cite{Stirling:1900sj}: due to shape assumptions, one
can find some regions with small errors despite the lack of data points. More
precisely, as shown by \cite{ref1yj}, uncertainties on the gluon distribution at
very low $x$ ($x \leq 10^{-4}$, where there are no constraining experimental 
data points)
obtained by CTEQ6.5 and MRST2001e do not overlap. More flexible shapes modeled
with neural networks and fitted using a genetic algorithm from NNPDF
\cite{Ball:2008by} give much larger error bands overlapping with the ones from 
CTEQ and
MRST. We note that dynamical PDFs generated radiatively from valence like input
at low scales may be another approach which yields smaller uncertainties, 
\cite{Gluck:2007ck}, see also fig. 3 of \cite{ref4yj}.

\begin{figure}[htb]
\begin{center}
\epsfig{file=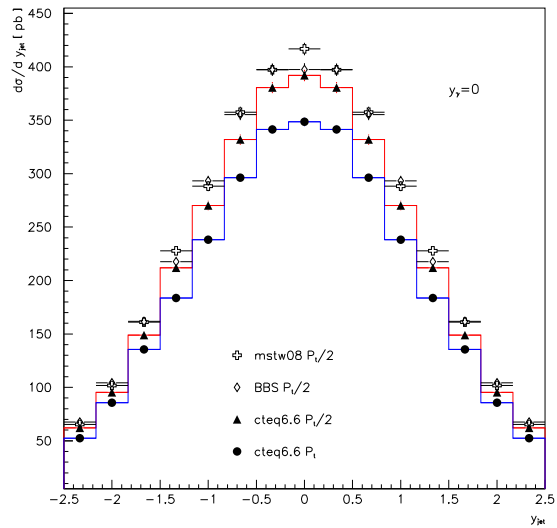, height=8cm}
\caption{\label{yjet2-00} 
Distribution of leading jet rapidity in photon + jet
associated production, at $y_{\gamma} =$ 0 for various pdf sets and scale 
choices.} 
\end{center}
\end{figure}

\begin{figure}[htb]
\begin{center}
\epsfig{file=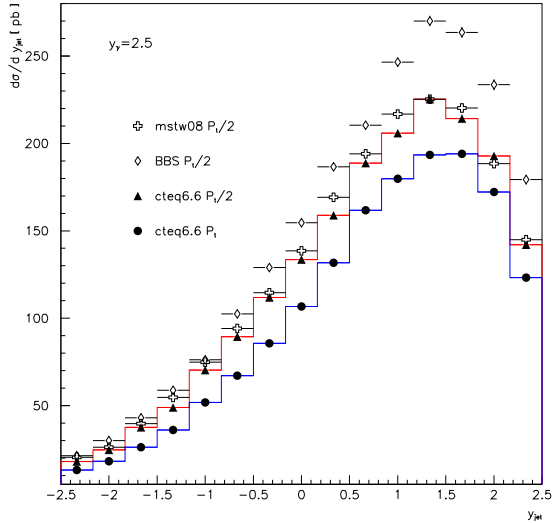, height=8cm}
\caption{\label{yjet2-25} 
Distribution of leading jet rapidity in photon + jet
associated production, at $y_{\gamma} =$ 2.5 for various pdf sets and scale 
choices.}
\end{center}
\end{figure}

Is it possible to be less sensitive to shape assumptions by using photon+jet 
correlations?
The Tevatron experiments CDF and D0 both during Runs I and II, have been 
measuring photon-jet correlations, which explore the short distance dynamics 
in a more constrained way than inclusive photon production. 
A recent comparison between D0 data and {\tt JETPHOX} has 
been performed \cite{Abazov:2008er} 
for the distribution $d \sigma/d y^{\gamma}$ vs. $p_{T}^{\gamma}$.
An interesting study\footnote{Yet we notice that this study accounts for the
fragmentation contribution at LO only.} \cite{Gupta:2008zza} of the possibilities of the CMS 
experiment on this distribution has recently appeared.
Among the other correlations which can be studied, let us mention 
the distribution of jet rapidity at fixed photon rapidity, integrated over the
photon transverse momenta above some $p_{T \, min}^{\gamma}$. 
At large rapidities, the main contribution comes from the subprocess 
$q g \to q \gamma$ (or $\bar{q} g\to \bar{q} \gamma$) where the initial state 
gluon is at quite 
low $x$, down to ${\cal O}($a few $10^{-3})$ while the $x$ of the initial state 
(anti)quark is $\sim {\cal O}(10^{-1})$. This correlation observable is thus
sensitive to gluons in a low $x$ region overlapping with the one explored at
HERA.

\vspace{0.3cm}

Figs. \ref{yjet2-00} and \ref{yjet2-25} show the {\tt JETPHOX} predictions for 
various pdf sets for the distribution of jet rapidity, 
for the photon rapidities $y_{\gamma} =$ 0 and 2.5, respectively 
at the Tevatron for $\sqrt{S} = 1.96$ TeV. The jets
are defined according to the D0 midpoint algorithm with cone aperture 
$R_{C} = 0.7$ \cite{jetD0}, and the discussion at NLO 
refers to the leading jet i.e. the jet with highest $p_{T}$. 
The cross section is integrated over photon transverse momenta larger than 
30 GeV and over jet $p_{T}$ larger than 20 GeV.  
The choice of scales is $\mu = M = M_{F} = p_{T}^{\gamma}/2$. 
Besides the prediction using\footnote{The photon fragmentation functions (PFF) 
are from BFG set II~\cite{3r}. Due to the D0 isolation requirement - less than 
2 GeV of accompanying hadronic transverse energy in a cone of radius 0.4 in 
azimut and rapidity around the photon direction - the prediction depends only
marginally on the PFF choice.} the pdf sets CTEQ 6.6 \cite{Nadolsky:2008zw}
and MSTW08 \cite{mstw08} resulting from global fits, we also show the 
prediction with the BBS set \cite{Bourrely:2001du}, an example of a set modeled 
through dynamical generation, which has a quite different gluon pdf in the low 
$x$ region, to illustrate the sensitivity of this observable to low $x$ gluons.
However we cannot draw any definite conclusions from this observable alone 
since the dependence of these predictions on the scale choice
at NLO is as large as the spread with respect to the various pdfs used.
An error analysis taking into account the detailed information provided by 
MSTW08 is beyond the scope of this paper.

\section{The Photon Fragmentation Function}\label{pho-frag}
\hspace*{\parindent}

The large-$p_{T}$ photon-jet correlations also give access to the Photon
Fragmentation Function (PFF) which has rarely been measured. Actually only two
LEP experiments, ALEPH \cite{1r} and OPAL \cite{2r} measured the PFF. However
it is difficult to observe a photon in  a large hadronic background and hence
the PFF has mainly been measured for large values of 
$z = 2E_{\gamma}/\sqrt{s}$. Good agreement is found between these data and two 
NLO theoretical results \cite{3r,4r}.

\begin{figure}[htb]
\begin{center}
\epsfig{file=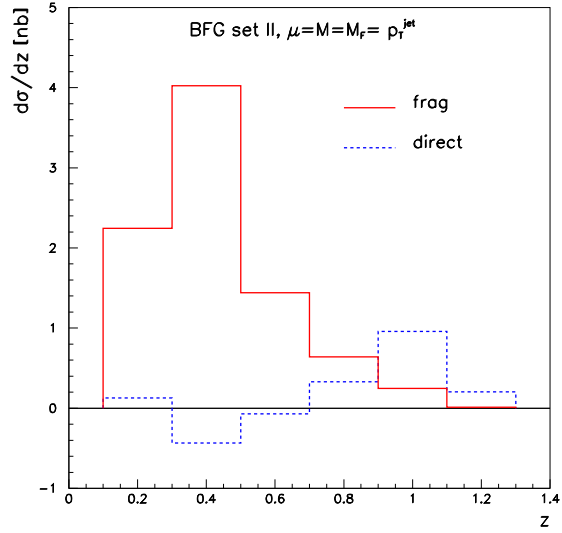, height=8cm}
\caption{\label{fig1} Cross section $d\sigma/dz_{\gamma}$ for the BFG set II.}
\end{center}
\end{figure}

\begin{figure}[htb]
\begin{center}
\epsfig{file=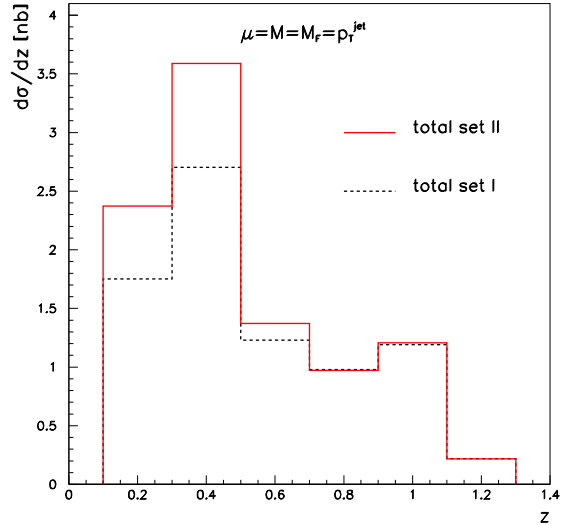, height=8cm}
\end{center}
\caption{\label{fig:set1set2} $d\sigma/dz_{\gamma}$ for the two BFG sets 
I and II.}
\end{figure}

\begin{figure}[htb]
\begin{center}
\epsfig{file=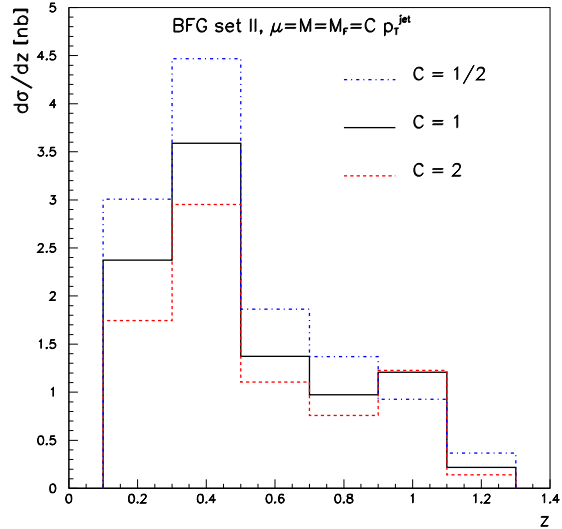, height=8cm}
\end{center}
\caption{\label{fig:scalesdiag} Scale dependence of $d\sigma/dz_{\gamma}$ for a
common variation of scale $\mu = M = M_{F}$.}
\end{figure}

\begin{figure}[htb]
\begin{center}
\epsfig{file=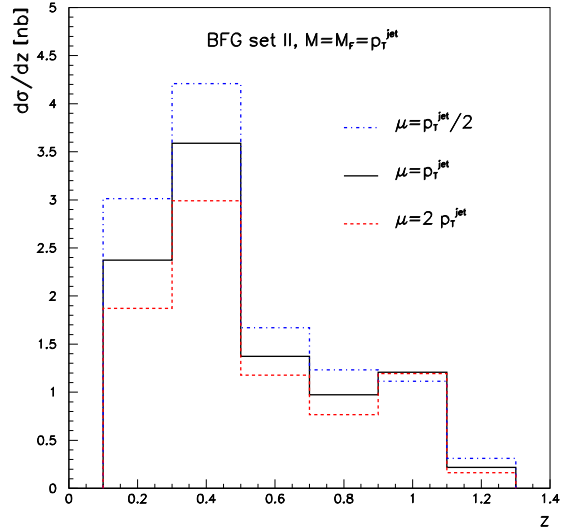, height=8cm}
\caption{\label{fig:murentot} Separate renormalization scale ($\mu$) 
dependence of $d\sigma/dz_{\gamma}$ for fixed common factorization scales 
$M = M_{F}=p_{T}^{\rm{jet}}$.}
\end{center}
\end{figure}

\begin{figure}[htb]
\begin{center}
\epsfig{file=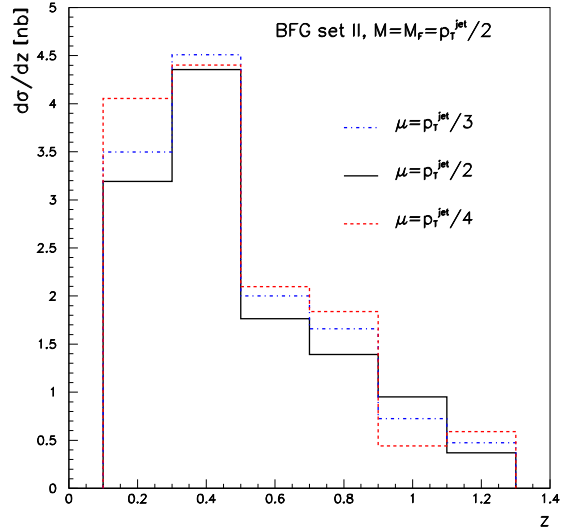, height=8cm}
\caption{\label{fig:opt} Region of larger stability w.r.t. renormalization
scale dependence.}
\end{center}
\end{figure}

\begin{figure}[htb]
\begin{center}
\epsfig{file=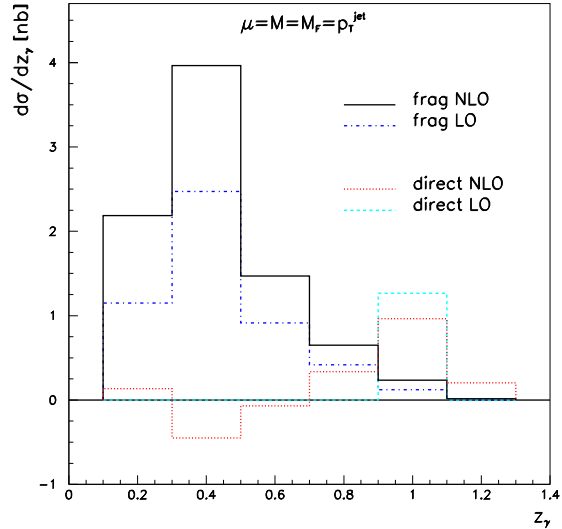, height=8cm}
\caption{\label{fig:ho} Respective sizes of HO corrections in fragmentation and 
direct components for a given common scale choice.}
\end{center}
\end{figure}

The hadroproduction of large-$p_{T}$ photons and jets should also allow to
measure the PFF. First let us consider the direct subprocess $qg \to q\gamma$ in
which the final photon transverse momentum $p _{T}^{\gamma}$ is balanced by the
$q$-jet transverse momentum $p_{T}^{\rm{jet}}$ such that 
$z_{\gamma} = - {\vec{p}_{T}^{\,\gamma} \cdot
\vec{p}_{T}^{\,\rm{jet}} \over |\!|\vec{p}_{T}^{\,\rm{jet}}|\!|^2} = 1$.  
The PFF is not involved in the description of this reaction.  The PFF manifests 
itself, for instance, in the subprocess reaction $gq \to gq$ followed by the 
gluon ($g \to \gamma + X$) or the quark ($q \to \gamma + X$) collinear 
fragmentations described by the distributions $D_{\gamma/g}(z_{\gamma}, M_F^2)$ 
and $D_{\gamma/q}(z_{\gamma}, M^2_F)$, the large scale $M^2_F$ being of order
$(p_{T}^{\gamma})^2$. At Leading Order (LO) we have $z_{\gamma} \leq 1$ and the
cross section $d\sigma^{\rm{frag}}/dz_{\gamma}$ is directly proportional to the
functions $D_{\gamma/a}(z_{\gamma}, M^2_F)$ ($ a = q, g$). Therefore at leading
order the total cross section $d\sigma/dz_{\gamma}$ is given by the sum of the
direct contribution (proportional to $\delta (1 -z_{\gamma})$) and of the
fragmentation contribution containing the PFF $D_{\gamma/a} (z_\gamma, M^2_F)$.
Contrarily to $e^+e^-$ experiments we have to include a direct contribution in
the cross section and to stay away from $z_{\gamma} = 1$ to increase the
sensitivity to the PFF.
\par

When HO corrections are taken into account, more jets can be  present in the
final state and $z_{\gamma}$ may be larger than one.  With three partons in the
final state, $z_{\gamma}$ can be different from one also in the direct
contribution.
\par

The variable $z_{\gamma}$ depends on the two transverse momenta 
$\vec{p}_{T}^{\,\gamma}$ and $\vec{p}_{T}^{\,\rm{jet}}$, none of them being a
priori fixed for a given value of $z_{\gamma}$. However $p_{T}^{\rm{jet}}$ is
directly related (at least at LO)  to the parton momentum involved in the hard
subprocess. Therefore, if we vary $p_{T}^{\rm{jet}}$ to obtain different values
of $z_{\gamma}$ (keeping $p_{T}^{\gamma}$ fixed), the theoretical cross section
$d\sigma/dz_{\gamma}$ will reflect the $z_{\gamma}$ dependence of
$D_{\gamma/a}(z_{\gamma}, M^2_F)$ {\bf and} the $p_{T}^{\rm{jet}}$ dependence of
the subprocess, thus blurring the $z_{\gamma}$-dependence of the PFF. On the
contrary, if we keep $p_{T}^{\rm{jet}}$ fixed and vary $p_{T}^{\gamma}$, we
obtain a $z_\gamma$-dependence of $d\sigma/dz_{\gamma}$ coming dominantly from
the fragmentation function.
Therefore we propose to measure the PFF in experiments in which the jet momentum
is kept fixed and the photon momentum is varied.
\par

Let us finally note that the observed photon must not be isolated, which would
considerably reduce the fragmentation contribution. This possibility exists 
when the photon-$p_{T}$ is not too large, as it is the case at RHIC for
$p_{T}^{\gamma} \ \lsim\
16$~GeV~\cite{Peressounko:2006qs,KleinBosing:2007bp,Jin:2007by,Frantz:2009zn}. 
Therefore we
choose the photon and jet momenta in agreement with the experiment performed at
RHIC, i.e. in our numerical analysis  we use  
$3 \leq p_{T}^{\gamma} \leq 16$~GeV, 11~GeV $\leq p_{T}^{\rm{jet}} \leq 13$~GeV 
and $\sqrt{s} = 200$~GeV.
For the rapidities we take $-{1 \over 2} \leq y^{\gamma} \leq {1 \over 2}$  and
$-1 \leq y^{\rm{jet}} \leq 1$.  We use the CTEQ6M parton distribution 
functions\,\cite{Pumplin:2002vw}, and renormalization and factorization scales
given by  $C\cdot p_{T}^{\rm{jet}}$ with ${1 \over 2} \leq C \leq 2$.

\subsection*{Discussion of numerical results}

In Fig.~\ref{fig1} we present the results for the cross section 
$d\sigma/dz_{\gamma}$, calculated with the scales $\mu=M=M_F=p_{T}^{\rm{jet}}$
and the set II of the BFG fragmentation functions~\cite{3r}. 

The jet is defined by the midpoint algorithm~\cite{jetD0} 
with $R_{\rm{cone}}=0.7$. 
We observe that at small values of $z_\gamma$, the fragmentation contribution 
is much larger than the direct one. Although 
at next-to-leading order, the fragmentation and direct contributions cannot 
be considered as independent physical channels, 
this feature persists for other scale 
choices.
Therefore the photon fragmentation functions should be measurable in RHIC 
experiments at small values of $z_{\gamma}$, a region which was not accessible 
to LEP experiments. 

\vspace{0.3cm}

In Fig.~\ref{fig:set1set2} we compare the predictions obtained with 
the sets I and II of the BFG fragmentation functions. 
The gluon fragmentation function of set I is smaller than the one of set II 
at small $z_{\gamma}$, leading to 
a significant difference in the $d\sigma/dz_{\gamma}$
distribution, as can be seen from  Fig.~\ref{fig:set1set2}. 
This result shows the sensitivity of this reaction to the 
photon fragmentation functions. 

\vspace{0.3cm}

We find that the scale dependence at NLO is about $\pm 20 \%$
as can be seen from Fig.~\ref{fig:scalesdiag}. 
There are three scales involved: the renormalisation scale 
$\mu$, the initial state factorisation scale $M$ 
(we set the factorisation scales for 
both initial state particles equal to $M$), 
and the final state factorisation scale $M_F$. 
The dependence of the cross section on these scales is quite different. 

\vspace{0.3cm}

In Fig.~\ref{fig:scalesdiag} we vary all three scales simultaneously.
If we vary the final state factorisation scale $M_F$
only, keeping $\mu=M$ fixed, 
the fragmentation component increases logarithmically 
with $M_F$, while the direct contribution decreases, 
such that the leading logarithmic dependence cancels in the sum
and the overall dependence on $M_F$ is rather weak. 
We also checked that the dependence on the initial state factorisation scale 
is very weak, such that  the overall scale uncertainty is dominated by the 
renormalisation scale dependence. This is demonstrated 
in Fig.~\ref{fig:murentot}, where we vary the renormalisation scale $\mu$
only, keeping $M=M_F$ fixed to $p_{T}^{\rm{jet}}$.
However, we can find a region where the cross section 
is more stable against variations of $\mu$, 
which is in the vicinity of $M=M_F =p_{T}^{\rm{jet}}/2$ and 
$p_{T}^{\rm{jet}}/4<\mu<p_{T}^{\rm{jet}}/2$, as can be seen from 
Fig.~\ref{fig:opt}. This optimal behavior is obtained in the 
small-$z$ domain in which the fragmentation contribution is large.

\vspace{0.3cm}

Fig.~\ref{fig:ho} shows that the size of the higher order corrections 
is much larger for the fragmentation component than for the direct component.

\section{Conclusions}\label{conclusions}

In this article we have studied photon-jet correlations in 
hadronic collisions based on the NLO program  {\tt JETPHOX}, 
which is a Monte Carlo program of partonic event generator type 
which incorporates NLO corrections to both direct photons and 
photons from fragmentation. The program is flexible to 
account for user-defined kinematic cuts and 
photon isolation parameters. It is available at the following web site 
\cite{http-jetphox}.
Correlation observables offer in general a larger sensitivity 
to the short distance dynamics than one-particle inclusive 
observables. 
We studied the photon-jet angular distribution $\cos\theta^*$
and the jet rapidity distribution in view of possible constraints on the 
parton distribution functions in the proton, in particular the gluon.

\vspace{3mm} 

Furthermore, correlations involving unisolated photons such as the ones
measured at RHIC provide a means to constrain the photon 
fragmentation function in a region which was barely accessible by the 
LEP experiments. 
We study the observable $d\sigma/dz_\gamma$, where $z_\gamma$ 
can be reconstructed from the photon and jet 
transverse momenta.
We argue that the  $z_\gamma$ dependence 
of $d\sigma/dz_\gamma$ is coming dominantly from the 
fragmentation if we keep $p_T^{\rm{jet}}$ fixed and vary $p_T^\gamma$,
and therefore propose to measure the fragmentation functions at fixed 
$p_T^{\rm{jet}}$.
The NLO predictions
still have a non negligible 
dependence on the renormalisation scale choice, 
which is related to the fact that the higher order corrections to the 
fragmentation component are large, exceeding 40\%. 

\vspace{0.3cm}

The program {\tt JETPHOX} can also predict the tail of the distribution of
transverse momentum $Q_{T}$ of photon+jet pairs at NLO accuracy for large 
enough $Q_{T}$ for the LHC, as well as any other correlation 
insensitive to multiple soft gluon emission, it thus can also help to
normalize the Monte Carlo event generators suited to describe the lower 
$Q_{T}$ range of the distribution of
transverse momentum of photon+jet pairs and other less inclusive observables.

\section*{Note added}
While we were completing the present article, we became aware of a work by 
Stavreva and Owens \cite{Stavreva:2009vi}, similar to the one reported here, 
but focusing on charm and bottom jets in the final state.


\begin{thebibliography}{99} 

%\cite{Dissertori:2005he}
%\bibitem{Dissertori:2005he}
%  G.~Dissertori,
  %``LHC expectations (machine, detectors and physics),''
%  PoS {\bf HEP2005} (2006) 401
%  [arXiv:hep-ex/0512007].
  %%CITATION = POSCI,HEP2005,401;%%

%\cite{Ball:2007zza}
\bibitem{Ball:2007zza}
  G.~L.~Bayatian {\it et al.}  [CMS Collaboration],
  %``CMS technical design report, volume II: Physics performance,''
  J.\ Phys.\ G {\bf 34} (2007) 995.
  %%CITATION = JPHGB,G34,995;%%
  
%\cite{:2008zzk}
\bibitem{:2008zzk}
  R.~Adolphi {\it et al.}  [CMS Collaboration],
  %``The CMS experiment at the CERN LHC,''
  JINST {\bf 0803}, S08004 (2008)
  [JINST {\bf 3}, S08004 (2008)].
  %%CITATION = JINST,3,S08004;%%

%\cite{Aad:2009wy}
\bibitem{Aad:2009wy}
  G.~Aad {\it et al.}  [The ATLAS Collaboration],
  %``Expected Performance of the ATLAS Experiment - Detector, Trigger and
  %Physics,''
  [arXiv:0901.0512].
  %%CITATION = ARXIV:0901.0512;%%

%\cite{:2008zzm}
\bibitem{:2008zzm}
  G.~Aad {\it et al.}  [ATLAS Collaboration],
  %``The ATLAS Experiment at the CERN Large Hadron Collider,''
  JINST {\bf 3}, S08003 (2008).
  %%CITATION = JINST,3,S08003;%%

\bibitem{Stockton:2008zz}
M.~Stockton  [ATLAS Collaboration],
%``Production of jets and photons at ATLAS,''
Nucl.\ Phys.\ Proc.\ Suppl.\  {\bf 186} (2009) 11.
%%CITATION = NUPHZ,186,11;%%

\bibitem{PerezReale:2008zza}
  V.~Perez-Reale  [ATLAS Collaboration],
  %``Physics With Photons At The Atlas Experiment,''
  Nucl.\ Phys.\ Proc.\ Suppl.\  {\bf 184}, 182 (2008).
  %%CITATION = NUPHZ,184,182;%%

%\cite{Buescher:2005re}
\bibitem{Buescher:2005re}
  V.~Buescher and K.~Jakobs,
  %``Higgs boson searches at hadron colliders,''
  Int.\ J.\ Mod.\ Phys.\  A {\bf 20} (2005) 2523
  [arXiv:hep-ph/0504099].
  %%CITATION = IMPAE,A20,2523;%%%\cite{Aurenche:2006vj}

%\cite{Stockli:2005hz}
\bibitem{Stockli:2005hz}
  F.~Stockli, A.~G.~Holzner and G.~Dissertori,
  %``Study of perturbative QCD predictions at next-to-leading order and  beyond
  %for p p --> H --> gamma gamma + X,''
  JHEP {\bf 0510} (2005) 079
  [arXiv:hep-ph/0509130].
  %%CITATION = JHEPA,0510,079;%%

\bibitem{Fang:2008zzc}
Estimates using {\tt JETPHOX} can be found in ref. \cite{Aad:2009wy}, 
table 1, p 1214, and in \cite{Stockton:2008zz}. See also:\\ 
Y.~Fang,
  %``Search for SM Higgs decaying to two photons via ATLAS detector,''
  %%CITATION = 
CERN-THESIS-2008-073, page 78, table 5.3. 

%\cite{Peressounko:2006qs}
\bibitem{Peressounko:2006qs}
  D.~Peressounko  [PHENIX Collaboration],
  %``Direct photon production in p + p and d + Au collisions measured with the
  %PHENIX experiment,''
  Nucl.\ Phys.\  A {\bf 783} (2007) 577
  [arXiv:hep-ex/0609037].
  %%CITATION = NUPHA,A783,577;%%

%\cite{Baer:1989xj}
\bibitem{Baer:1989xj}
  H.~Baer, J.~Ohnemus and J.~F.~Owens,
  %``A Calculation Of The Direct Photon Plus Jet Cross-Section In The
  %Next-To-Leading Logarithm Approximation,''
  Phys.\ Lett.\  B {\bf 234} (1990) 127.
  %%CITATION = PHLTA,B234,127;%%

\bibitem{Aurenche:2006vj}
P.~Aurenche, M.~Fontannaz, J.~P.~Guillet, E.~Pilon and M.~Werlen,
%``A new critical study of photon production in hadronic collisions,''
Phys.\ Rev.\  D {\bf 73} (2006) 094007
[arXiv:hep-ph/0602133].

\bibitem{http-jetphox}
Code available at the following URL address:\\
{\tt http://lappweb.in2p3.fr/lapth/PHOX\_FAMILY/main.html}

\bibitem{Abe:1993cv}
  F.~Abe {\it et al.}  [CDF Collaboration],
  %``The Center-of-mass angular distribution of prompt photons produced in
  %$p\bar{p}$ collisions at $\sqrt{s} = 1.8$ TeV,''
  Phys.\ Rev.\ Lett.\  {\bf 71} (1993) 679.
  %%CITATION = PRLTA,71,679;%%

 %\cite{Lamoureux:1995ps}
\bibitem{Lamoureux:1995ps}
  J.~I.~Lamoureux,
  %``Photon production at CDF and D0,''
  AIP Conf.\ Proc.\  {\bf 357}, 548 (1996).
  %%CITATION = APCPC,357,548;%%

\bibitem{cfgp}
S.~Catani, M.~Fontannaz, J.P.~Guillet and E.~Pilon, 
JHEP 0205 (2002) 28
[arXiv:hep-ph/0204023].

%\cite{Abe:1993qb}
\bibitem{Abe:1993qb}
  F.~Abe {\it et al.}  [CDF Collaboration],
  %``A Prompt photon cross-section measurement in $\bar{p}p$ collisions at
  %$\sqrt{s} = 1.8$ TeV,''
  Phys.\ Rev.\  D {\bf 48} (1993) 2998.
  %%CITATION = PHRVA,D48,2998;%%

%\bibitem{diphoton}
\bibitem{Acosta:2004sn}
  D.~E.~Acosta {\it et al.}  [CDF Collaboration],
  %``Measurement of the cross section for prompt diphoton production in
  %$p\bar{p}$ collisions at $\sqrt{s} = 1.96$ TeV,''
  Phys.\ Rev.\ Lett.\  {\bf 95} (2005) 022003
  [arXiv:hep-ex/0412050].
  %%CITATION = PRLTA,95,022003;%%

%\cite{Stirling:1900sj}
\bibitem{Stirling:1900sj}
  W.~J.~Stirling,
  %``Progress in Parton Distribution Functions,''
  [arXiv:0812.2341 [hep-ph]].
  %%CITATION = ARXIV:0812.2341;%%

\bibitem{ref1yj}
 Harrison Prosper, invited talk at ACAT 2008, Erice,
%{\tt http://indico.cern.ch/contributionDisplay.py?contribId=229&sessionId=11&confId=34666}\\
to appear in POS

%\cite{Ball:2008by}
\bibitem{Ball:2008by}
  R.~D.~Ball {\it et al.}  [NNPDF Collaboration],
  %``A determination of parton distributions with faithful uncertainty
  %estimation,''
  Nucl.\ Phys.\  B {\bf 809} (2009) 1
  [arXiv:0808.1231 [hep-ph]].
  %%CITATION = NUPHA,B809,1;%%

%\cite{Gluck:2007ck}
\bibitem{Gluck:2007ck}
  M.~Gluck, P.~Jimenez-Delgado and E.~Reya,
%  %``Dynamical parton distributions of the nucleon and very small-x physics,''
  Eur.\ Phys.\ J.\  C {\bf 53} (2008) 355
  [arXiv:0709.0614 [hep-ph]].
  %%CITATION = EPHJA,C53,355;%%

\bibitem{ref4yj}
 J.~Bl\"umlein, talk at DIS2007 [arXiv:0711.1721v1]

%\cite{Abazov:2008er}
\bibitem{Abazov:2008er}
  V.~M.~Abazov {\it et al.}  [D0 Collaboration],
  %``Measurement of the differential cross-section for the production of an
  %isolated photon with associated jet in $p \bar{p}$ collisions at $\sqrt{s}$ =
  %1.96-TeV,''
  Phys.\ Lett.\  B {\bf 666} (2008) 435
  [arXiv:0804.1107 [hep-ex]].
  %%CITATION = PHLTA,B666,435;%%

%\cite{Gupta:2008zza}
\bibitem{Gupta:2008zza}
  P.~Gupta, B.~C.~Choudhary, S.~Chatterji and S.~Bhattacharya,
  %``Study of direct photon plus jet production in CMS experiment at $\sqrt{s}$
  %= 14-TeV,''
  Eur.\ Phys.\ J.\  C {\bf 53} (2008) 49.
  %%CITATION = EPHJA,C53,49;%%

\bibitem{jetD0}
%QCD and Weak Boson Physics in Run II, FERMILAB-PUB-00/297, p. 62-63
G.C~Blazey et al, 
%Run II Jet Physics, 
[arXiv: hep-ex/0005012v2],
%Run II Proceedings of the Run II QCD and 
published in {\em Batavia 1999, Weak Boson Physics in Run II}, 
FERMILAB-PUB-00/297, 47-77, edited by. U.~Baur, R.~K.~Ellis and D.~Zeppenfeld.
 
%\cite{Nadolsky:2008zw}
\bibitem{Nadolsky:2008zw}
  P.~M.~Nadolsky {\it et al.},
  %``Implications of CTEQ global analysis for collider observables,''
  Phys.\ Rev.\  D {\bf 78}, 013004 (2008)
  [arXiv:0802.0007 [hep-ph]].
  %%CITATION = PHRVA,D78,013004;%%


\bibitem{mstw08}
A. D. Martin, W. J. Stirling, R. S. Thorne and G. Watt,
[arXiv:0901.0002 [hep-ph]]


%\cite{Bourrely:2001du}
\bibitem{Bourrely:2001du}
  C.~Bourrely, J.~Soffer and F.~Buccella,
  %``A statistical approach for polarized parton distributions,''
  Eur.\ Phys.\ J.\  C {\bf 23} (2002) 487
  [arXiv:hep-ph/0109160].
  %%CITATION = EPHJA,C23,487;%%

\bibitem{3r}
L. Bourhis, M. Fontannaz and J. Ph. Guillet, 
Eur. Phys. J.  {\bf C2} (1998) 529. 

\bibitem{1r} 
ALEPH collaboration, D. Buskulic et al., 
Z. Phys.  {\bf C69} (1996) 365.

\bibitem{2r} 
OPAL collaboration, K. Ackerstaff et al., 
Eur. Phys. J.  {\bf 2} (1998) 39.

\bibitem{4r}  
M. Gl\"uck, E. Reya and A. Vogt, 
Phys. Rev. {\bf D48} (1993) 116.
%\cite{Jin:2007by}
  
%\cite{KleinBosing:2007bp}
\bibitem{KleinBosing:2007bp}
  C.~Klein-Bosing  [PHENIX Collaboration],
  %``Systematic study of high-$p_T$ hadron and photon production with the PHENIX
  %experiment,''
  [arXiv:0710.2960 [nucl-ex]].
  %%CITATION = ARXIV:0710.2960;%%
  
\bibitem{Jin:2007by}
  J.~Jin  [PHENIX Collaboration],
  %``PHENIX measurement of high-p(T) hadron hadron and photon hadron azimuthal
  %correlations,''
  J.\ Phys.\ G {\bf 34} (2007) S813
  [arXiv:0705.0842 [nucl-ex]].
  %%CITATION = JPHGB,G34,S813;%%

%\cite{Frantz:2009zn}
\bibitem{Frantz:2009zn}
  J.~Frantz,
  %``Two-particle Direct Photon-Jet Correlation Measurements in PHENIX,''
  arXiv:0901.1393 [nucl-ex].
  %%CITATION = ARXIV:0901.1393;%%

\bibitem{Pumplin:2002vw}
  J.~Pumplin, D.~R.~Stump, J.~Huston, H.~L.~Lai, P.~Nadolsky and W.~K.~Tung,
  %``New generation of parton distributions with uncertainties from global  QCD
  %analysis,''
  JHEP {\bf 0207} (2002) 012
  [arXiv:hep-ph/0201195].
  %%CITATION = JHEPA,0207,012;%%

%\cite{Stavreva:2009vi}
\bibitem{Stavreva:2009vi}
  T.~P.~Stavreva and J.~F.~Owens,
  %``Direct Photon Production in Association With A Heavy Quark At Hadron
  %Colliders,''
  [arXiv:0901.3791 [hep-ph]].
  %%CITATION = ARXIV:0901.3791;%%

\end{thebibliography}
\end{document}